\documentclass[12pt]{spieman}  

\usepackage{amsmath,amsfonts,amssymb}
\usepackage{rotating}
\usepackage{hyperref}
\usepackage{setspace}
\usepackage{tocloft}
\usepackage{enumitem}
\usepackage{ltablex}
\usepackage{graphicx}
\usepackage{float}
\usepackage{threeparttable}
\usepackage{hhline}
\usepackage{lineno}
\usepackage[normalem]{ulem}
\title{Assessing the Suitability of H4RG Near Infrared Detectors for Precise Doppler Radial Velocity Measurements} 

\author[]{Eric B. Bechter}
\author[]{Andrew J. Bechter}
\author[]{Justin R. Crepp}
\author[]{Jonathan Crass}
\affil[]{University of Notre Dame, Department of Physics, Nieuwland Science Hall, Notre Dame, IN 46556, USA}

\cftpagenumbersoff{figure}
\cftpagenumbersoff{table} 
\begin{document} 
\maketitle

\begin{abstract}
At wavelengths longwards of the sensitivity of silicon, hybrid structured mercury-cadmium-telluride (HgCdTe) detectors show promise to enable extremely precise radial velocity (RV) measurements of late-type stars. The most advanced near infrared (NIR) detector commercially available is the HAWAII series (HxRG) of NIR detectors. While the quantum efficiency of such devices has been shown to be $\approx 90\%$, the noise characteristics of these devices, and how they relate to RV measurements, have yet to be quantified. We characterize the various noise sources generated by H4RG arrays using numerical simulations. We present recent results using our end-to-end spectrograph simulator in combination with the “HxRG Noise Generator,” which emulates the effects of read noise, parameterized by white noise, correlated and uncorrelated pink ($1/f$) noise, alternating column noise, and picture frame noise. The effects of nonlinear pixel response, dark current, persistence, and interpixel capacitance (IPC) on RV precision are also considered. Our results have implications for RV error budgets and instrument noise floors that can be achieved with NIR Doppler spectrographs that utilize this kind of detector. 

\end{abstract}

{\noindent \footnotesize\textbf{Address all correspondence to:} E. Bechter  
\\ Email:\linkable{ebechter@nd.edu}\\ Intended for \textbf{JATIS} }

\begin{spacing}{1.5}   

\section{Introduction}
\label{sect:intro}
Astronomical observations in the near-infrared (NIR) offer sensitivity to faint M-dwarfs, which are known to host multiplanet systems in compact orbital configurations, while also providing lower inherent susceptibility to stellar variability as a result of lower contrast between surface inhomogeneities and the stellar photosphere \cite{Dressing_15,Marchwinski_15,Prato_15}. As such, a number of precise Doppler spectrographs are being developed to target this spectral region ($\lambda= 0.9-2.5~\mu$m), including systems that use adaptive optics to correct for atmospheric seeing \cite{Mahadevan_12,Thibault_12,Quirrenbach_16,Crepp_14,Jovanovic_16,Crepp_16}. At the same time, ever-increasing demands for wavelength coverage and spectral resolution require large format, two-dimensional arrays to offer sufficient pixels to sample dispersed starlight.

An apparent disadvantage to operating in the NIR beyond the cut-off of silicon is the required use of infrared detectors, which historically have lagged behind CCDs in terms of number of pixels and noise characteristics. However, recent advances in detector technologies have resulted in NIR arrays that are competitive in both categories with CCDs. The best NIR detectors available are the Hawaii-xRG (HxRG) family of detectors, where the `x' denotes the square array dimensions in units of thousands of pixels. HxRG arrays are based on a 2-layer structure consisting of a layer of mercury-cadmium-telluride (HgCdTe) and a layer of silicon that processes the charge. The two layers are combined using indium interconnects (one indium bond per pixel) creating a hybrid structure that takes advantage of the light sensing advantages of the HgCdTe layer and the readout technology of the CMOS silicon circuitry. The top layer of HgCdTe is specifically designed to be sensitive to photons at NIR wavelengths. The ratio of Hg to Cd of the HgCdTe alloy is responsible for the long cutoff wavelength, which can be manufactured at a number of values, e.g. $\mathrm{\lambda_{cutoff} = 1.7,~2.5,~and~5\mu m}$, depending on the science goals of the instrument. The silicon grid that makes up the lower layer is used as a read out integrated circuit (ROIC). Each pixel has all the necessary parts to amplify, multiplex, and extract the signals allowing for individual pixel reads and resets.

RVs computed from typical echelle format Doppler spectrographs are a statistical measure of typically hundreds of absorption line shifts all contained in a single detector exposure. For an R=100,000 resolving power instrument operating at 1 micron with 3 pixel sampling, each pixel is represented by about $\Delta{\rm RV}=1000$ ms$^{-1}$, which requires a sub-pixel absorption line tracking and centroiding precision of $<10^{-3}$ in order achieve RV precisions of less than $1~\mathrm{ms}^{-1}$ . There is already some precedence of NIR detectors being used at this level, although only tentatively in the precision RV field. The visible and NIR spectrograph, CARMENES, implements a mosaic of two H2RG detectors for the NIR arm. Having achieved $5-10~\mathrm{ms}^{-1}$ shortly after first light, further improvements are expected with additional development of calibration techniques\cite{Quirrenbach_18}. SPIRou, a new high resolution echelle spectrograph for the Canada-France-Hawaii Telescope employs a new H4RG with lab demonstrations indicating 0.2~ms$^{-1}$ stability over 24 hours and an expected on-sky precision of $1~\mathrm{ms}^{-1}$. However, outside of the RV field, astrometric measurements using NIRC2 and adaptive optics on the Keck II telescope have already been used to demonstrate $10^{-3}$ pixel precision imaging since at least 2005\cite{Ghez_08,Wizinowich_00}. NIRC2 utilizes an Aladdin III array produced by Raytheon, which differs from HxRGs by the use of indium antimonide (InSb) as a photosensitive material. Despite these material differences, successful imaging demonstrations indicate that NIR hybrid arrays will permit sub-meter-per-second Doppler precision on large telescopes.

The aim of this study is to connect H4RG detector noise characteristics directly to RV uncertainties in order to asses the suitability of H4RGs for use in precision Doppler spectrographs. Additionally, the findings of this paper have already proved to be an invaluable resource for conceptual, preliminary and final instrument design review stages. Our simulations consist of a combination of previously developed simulation code for the iLocater spectrograph with a Python-based HxRG detector noise generator\cite{BechterE_19, Rauscher_15}. Section \ref{sec:method} describes the general methodology behind our simulations and a description of how detector noise is translated into RV uncertainties. Section \ref{sec:noise} provides theory and practical models implemented for each noise source. Section \ref{sec:results} presents results for each detector effect along with a discussion of possible mitigation strategies presented in the literature. Section \ref{sec:concl} provides a summary of results, discussion of limitations, and further work to be done.

\section{Methodology}
\label{sec:method}
In this section, we illustrate the general methodology used to connect simulated detector effects to instrumental RV precision. Our analysis utilizes numerical simulation code developed to investigate designs for the \emph{iLocater} spectrograph, an NIR instrument being deployed at the Large Binocular Telescope on Mt. Graham.\cite{Crepp_14}. Unless otherwise noted, we use specifications from this instrument (bandpass, resolution, sampling, cross-dispersion) as a representative test case for evaluating the contributions from detector noise and its impact on RV precision. The fundamentals of the code operate by importing an optical model from Zemax/OpticStudio to \textsc{Matlab}, constructing a wavelength-dependent throughput budget for the instrument and mapping the input spectrum onto a two-dimensional grid according to the dispersion solution and instrument point-spread-function\cite{BechterA_18, BechterE_19}. The simulation package includes a catalog of synthetic stellar models used to propagate a spectrum of starlight through the system, obtained from Allard et al. 2012\cite{Allard_2012}.

\begin{figure*}
    \centering
    \includegraphics[width=\textwidth]{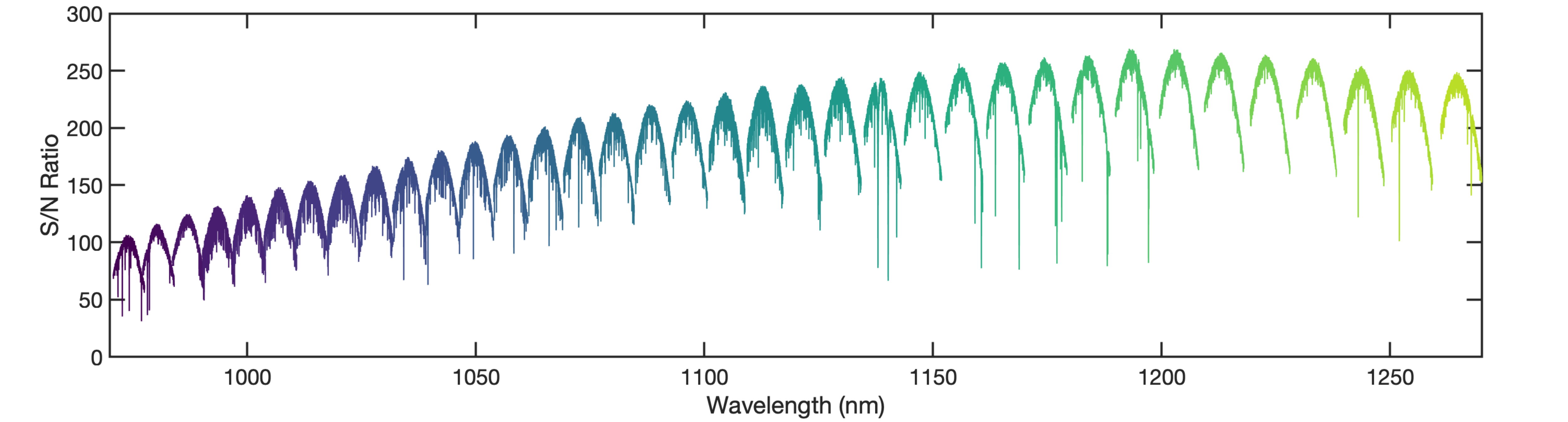}
    \caption{Simulation of an M0V, $I=10$ Echelle spectrum used to recover RV precision associated with several different sources of noise. A total of 36 spectral traces span grating orders $m = 117-152$. Contributions from tellurics, sky background, instrumental instabilities, or other contaminants are not included to better isolate individual detector effects under consideration.}
    \label{fig:input_spectrum}
\end{figure*}

To quantify RV noise caused by detector effects, a consistent spectral signal is used to recover Doppler information. Unless otherwise noted, an M0V with apparent magnitude $I = 10$ synthetic stellar spectrum was chosen for all simulations and tests. 
The simulated detector image is integrated to have a median photon-noise-dominated signal-to-noise ratio of $S/N = 187$ per collapsed pixel across all orders. Here, ``collapsed pixel'' describes the data reduction step of recovering a spectral trace and collapsing it in the cross-dispersion direction, resulting in a 1D vector. Figure~\ref{fig:input_spectrum} shows the result of the simulated star, with the y-axis indicating photon-noise limited S/N. iLocater's data reduction pipeline is used to locate and extract spectral orders, apply a wavelength solution and cross-correlate extracted spectra. A custom built binary mask is used to derive wavelength shifts in extracted spectra and compute the radial velocity offset. This technique has been used previously to build a comprehensive RV error budget for iLocater\cite{BechterA_18}.

In summary, H4RG noise is translated into RV precision using the following methodology: 
\begin{enumerate} 
\item[1.] Create a noise-free simulation of a stellar signal imprinted on a detector grid; 
\item[2.] Model the specific noise source that either generates an independent noise frame or modifies the signal frame according to models described in \S~\ref{sec:noise};  
\item[3.] Repeat 2. for a range of noise values and combine signal frame with noise frame if necessary;
\item[4.] Extract spectrum from simulated detector scene and recover RV with pipeline; 
\item[5.] Subtract mask offset values to obtain residual RV error. 
\end{enumerate} 

\section{Noise Models}
\label{sec:noise}
\subsection{Read Noise}
\begin{figure*}
\centering
\includegraphics[width=\textwidth]{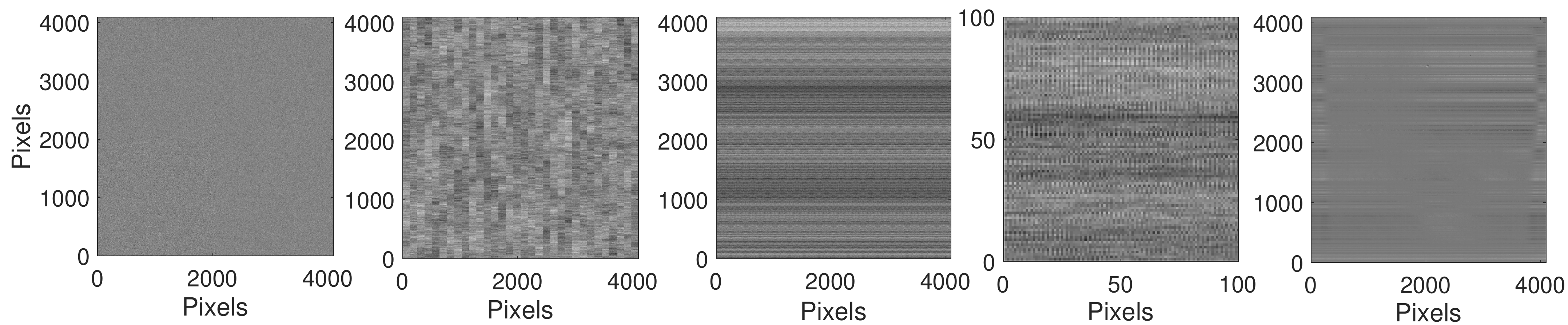}
\caption{PCA components of the noise generator. From left to right, panels show white noise, uncorrelated pink noise, correlated pink noise, alternating column noise and picture frame noise. Alternating column noise is shown as a sub-frame to highlight the characteristic high spatial frequency variations while the rest of the frames are full 4k$\times$4k frames.}
\label{fig:RN}
\end{figure*}

Read noise is added by making use of a noise generator (NG) originally developed to explore new readout modes for the James Webb Space Telescope Near Infrared Spectrograph (NIRSpec).\cite{Rauscher_13} The NG is based on principal component analysis (PCA) studies performed on the power spectrum of NIRSpec H2RGs. Noise sources include uncorrelated, correlated, stationary, and non-stationary components. Furthermore, the code separates noise components into the following categories: white noise; correlated pink noise; uncorrelated pink noise; alternating column noise; and picture frame noise. Figure~\ref{fig:RN} shows representative images of each PCA component.  

White noise appears as a uniform low power noise floor at relatively high spatial frequencies across the detector frame, originating primarily from shot noise due to leakage currents and charge traps, and Johnson noise from pixel interconnect resistance. Pink noise describes the characteristic $1/f$ slope in the detector's power spectrum and is comprised of both correlated and uncorrelated components. Physically, correlated pink noise appears as a horizontal banding that spans the entire detector width and uncorrelated pink noise results in a similar banding effect but is contained within a particular video output channel. 

Alternating column noise (ACN) manifests as high frequency, high power noise. Physically, ACN comes from the implementation of column buses resulting in even and odd detector columns carrying some $1/f$ noise, similar to uncorrelated pink noise but existing at a higher spatial frequency\cite{Rauscher_15, Rauscher_17}. Picture frame noise is perhaps the most visually unique noise pattern of the five noise components described. It appears as a framing component around the edges of the detector that vacillates in and out of exposures. It is believed that detector temperature fluctuations at the milli-Kelvin level can affect the magnitude of the framing component and is highly correlated from one frame to another\cite{Regan_08,Robberto_14}. 

The output of NG replicates the remaining noise pattern after standard reference pixel subtraction and up-the-ramp sampling and correction. NG simulations used in our analysis use all of the above components set to typical H4RG values. For the required zeroth principal component input we use the frame provided in NG's software download, a frame containing the ``picture frame'' components of a NIRspec H2RG. While NG is fundamentally based on H2RGs, it also provides settings to more closely resemble newer H4RGs, including updated noise values, a 4k$\times$4k array with H4RG output channel options and additional reference pixels.

\subsection{Pixel Linearity}
\label{sec:nonlinear}
The nonlinear pixel response of HxRG detectors is well studied in the literature\cite{Plazas_17, Biesiadzinski_11,Artigau_18}. The intrinsic nonlinear pixel response originates from several steps along the signal path. The first, reciprocity failure, describes the change in charge accumulation as a function of photon rate. This effect dominates at low flux levels. At higher flux, the charge-voltage conversion becomes the dominant term. Additionally, some elements along the electronic patch, e.g. source followers in the readout integrated circuit (ROIC) are also responsible for nonlinear effects.\cite{Plazas_17}

\begin{figure}[ht]\label{fig:linearitymodel}
\begin{center}
\includegraphics[width=0.8\textwidth]{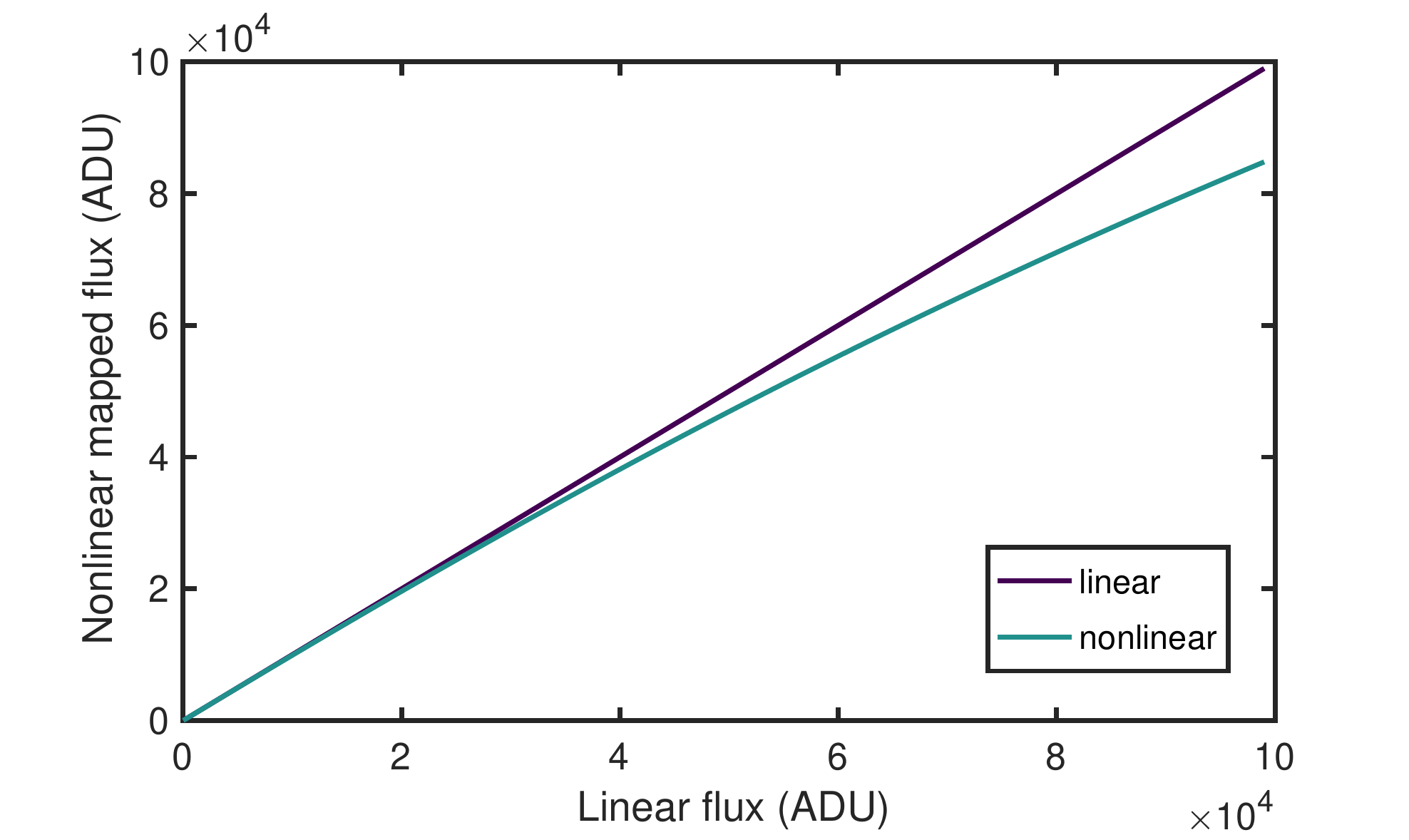}
\end{center}
\caption{Mapping between linear flux and nonlinear flux. The nonlinear curve shown in blue is derived in Artigau. For reference, the deviation from linearity at 30000 ADU is approximately 3\%.\cite{Artigau_18}}
\end{figure}

To simulate this nonlinear response, an empirical mapping between nonlinear and linearized flux, measured by Artigau,\cite{Artigau_18} is implemented. In their work, a correction term is fit to a few randomly selected pixel responses. Their correction term follows a cubic polynomial: 

\begin{equation}
   f_{corr} = f_{meas} + c_2f^2_{meas} + c_3f^3_{meas},
\label{eq:nonlincorrected}
\end{equation}

where $f_{meas}$ is the measured flux of a pixel and $f_{corr}$ is the linearized flux. They find median values of coefficients $c_2$ and $c_3$ to be $1.3\times10^{-6}$ and $1.4\times10^{-12}$, respectively. We invert this mapping function and apply it to simulated frames to reproduce the nonlinear effect at a variety of integration times. Figure \ref{fig:linearitymodel} shows the linear and nonlinear response curve used in our simulations. All pixels are assumed to respond according to the same model as it has been shown there is little variation between pixels.\cite{Artigau_18}

\subsection{Dark Current}
The manifestation of dark current in H4RGs is very similar to that in traditional CCDs. Dark current is typically described as the sum of several contributing physical effects including: diffusion, generation-recombination, tunneling, and surface current.\cite{bacon_06} Of these, diffusion and generation-recombination currents are highly dependent on temperature and can be reduced by cooling the detector, while the remaining effects largely result from manufacturing defects\cite{McLean_08}. HgCdTe detectors will generally produce constant dark current levels below a specific threshold temperature. However, beyond that threshold, the dark current sharply increases. In the case of H4RG-15 $1.5\mu m$ cutoff detectors, this threshold is around 100K and for temperatures exceeding 130K, every additional 10K results in an order of magnitude increase\cite{Zandian_16}. Therefore, instruments aiming for very low detector noise must target operating temperatures at or below the threshold temperature.

The appearance of dark current typically results in a fixed-pattern on the image frame. Examples of dark current frames in HxRGs are can be found in several detector studies\cite{Zandian_16,Piquette_14,Hall_16}. Figure~4 in Hall et al. 2016 shows a that the regions of particularly high dark current are generally isolated to the edges of the detector. During image processing, dark frame subtraction can remove an estimate of the mean or median fixed pattern; however residual noise remains due to the associated Poisson noise. In this study, residual dark current noise is simulated by randomly drawing from a Poisson distribution for each pixel in the detector and subtracting the mean value, resulting in a zero-centered, residual dark current frame. As dark current depends on operating temperatures, literature values for several different detector systems and different operating temperatures are simulated (\S~\ref{sec:darkresults}). It is also important to normalize literature values reported for detectors with different pixel dimensions to the same value\cite{Beletic_08}. In this study, we adopt $10\mu m$ pixel pitch as the default. As dark current is accumulated over time, $\Delta t = 30$ minutes is used, in agreement with iLocater's nominal integration time.

\subsection{Persistence}

The term persistence in relation to HgCdTe detectors refers to an increase in dark current after illumination. This causes a residual image of the previous exposure to remain in following images that decays over time\cite{Smith_08,Anderson_14,Mcleod_16}. Detector resets fail to clear the remnant image and removing persistence during image reduction proves challenging. 
Image persistence can adversely affect a number of instrument processes including target scheduling, calibration timing, data reduction and analysis procedures, and instrument measurement precision, all of which negatively impact survey science goals and efficiency. For example, during nightly observations, an instrument that surveys targets without allowing for detector persistence to reduce to zero will contain contaminating signals from previous targets. If the targets happen to include successive visits to stars with similar spectral types and absolute RVs, this will induce an RV shift that will add significantly to the error budget.

Smith\cite{Smith_08} developed a model for persistence based on a charge capture and emission process. Following from this work, Anderson\cite{Anderson_14} describes a 3D trap map. For our simulations, we adopt a simple model similar to that used in Artigau\cite{Artigau_18}. This model proposes that the level of persistence at the beginning of a new exposure is proportional to the accumulated photons at the time the pixel is reset. Following the previously mentioned work done by Artigau, persistence levels are assumed to decay according to $1/t$. For our purposes a one minute delay is used between the end of one exposure and the beginning of the next. Thus, to simulate persistence, a fractional persistent signal is applied to the primary stellar signal. Varying levels of persistence are investigated consistent with the persistence constant derived in Artigau. In addition, combinations of heterogeneous and homogeneous parasitic sources are investigated, each at a wide range of RV offsets. 


\subsection{Interpixel Capacitance}
\label{sec:ipc_theory}
Interpixel capacitance (IPC) describes electrical capacitive coupling between detector pixels. In theory, every pixel is coupled to every other pixel, not just nearest neighbors. Unlike charge diffusion in CCDs, where elements of charge drift from one pixel to another, IPC originates from fringing fields\cite{Kannawadi_16}. As the H4RG pixel size is manufactured smaller than H2RGs (10 or 15 $\mu$m compared to 18 $\mu$m), distance between neighboring pixels is reduced and IPC increases, becoming an important effect\cite{Finger_06}. IPC impacts RV precision by modifying point-spread-functions, potentially reducing spectral resolving power and disrupting absorption line profile shapes.

The effects of IPC can be simulated through the convolution of a kernel with the detector frame. IPC has a fixed pixel scale and has no dependence on wavelength, making it straightforward to simulate by convolving the simulated detector frame with an IPC kernel.  The simplest IPC kernel is modeled using a normalized discrete, 3$\times$3 square kernel\cite{Kannawadi_16}: 
\begin{equation}
   K(i,j) = \begin{pmatrix} 
      0 & \alpha & 0  \\ 
      \alpha & 1-4\alpha & \alpha  \\ 
      0 & \alpha & 0,
      \end{pmatrix}
\label{eq:simpleIPC}
\end{equation}
where $K(i,j)$ is the IPC kernel for pixel $i,j$ and $\alpha$ is the coupling fraction. IPC kernels are also known to have non-zero off diagonal components\cite{Hilbert_11}. Equation \ref{eq:simpleIPC} can be modified to reflect this:  
\begin{equation}
   K'(i,j) = \begin{pmatrix} 
      \alpha' & \alpha & \alpha'  \\ 
      \alpha & 1-4(\alpha+\alpha') & \alpha  \\ 
      \alpha' & \alpha & \alpha' 
      \end{pmatrix}.
\label{eq:diagIPC}
\end{equation}
Several values for $\alpha$ and $\alpha'$ are explored using this model including those expected for WFIRST's H4RG-10 detectors. Additionally, IPC kernels reported by SPIRou and Hubble's WFC3 are investigated\cite{Kannawadi_16,Artigau_18}.

\section{Simulation Results}
\label{sec:results}

\subsection{Read Noise}

We investigate the impact of read noise on RV precision by combining the residual read noise frame with the spectral signal from Figure~\ref{fig:input_spectrum} at various signal-to-noise values. Analysis of the simulated residual noise frame from the NG yields an rms of 5.4 e-, comparable to other values reported in the literature\cite{Artigau_18}. For each combined signal and noise frame, the RV error is assessed using masked cross-correlation, including subtracting the offset RV of the weighted binary mask that isolates stellar absorption lines. 

\begin{figure}[ht]\label{fig:RNresult}
\begin{center}
\includegraphics[width=0.7\textwidth]{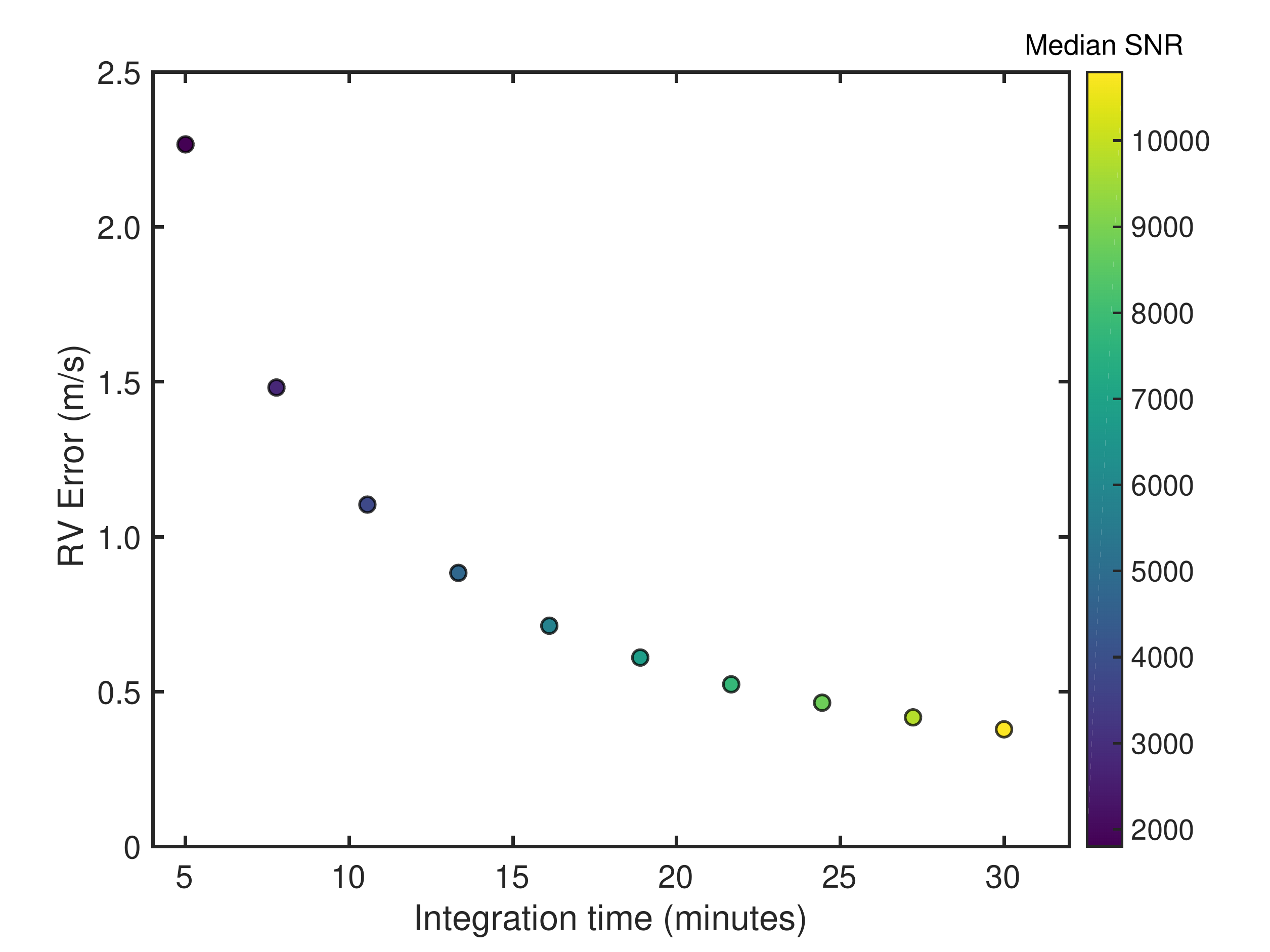}
\end{center}
\caption{The effect of residual read noise on RV precision as a function of integration time. Colors indicate median signal-to-noise ratios for each sample.}
\end{figure} 

The impact of read noise at varying integration times is shown in Figure~\ref{fig:RNresult}. For the nominal simulation parameters described in $\S$~\ref{sec:method}, an integration time of $\Delta t=30$ minutes corresponds to an RV uncertainty of $\sigma_{RV}=37~\mathrm{cms}^{-1}$. For RV instruments attempting sub-ms$^{-1}$ precision, an error this large is likely problematic as it leaves no room for remaining effects in the error budget. For reference, iLocater’s error budget has allocated a total of 30~cms$^{-1}$ to all detector effects in order to achieve an overall single measurement precision of 75 cms$^{-1}$ on M0, $I=10$ magnitude stars.\cite{BechterA_18}  

Further reducing HxRG read noise will be crucial to provide sufficient room for other error terms in precision RV budgets. Improved Reference Sampling and Subtraction (IRS2), developed by Bernard Rauscher at NASA GSFC, is a newly developed readout technique to reduce read noise in near infrared detectors\cite{Rauscher_17}. Originally developed for JWST H2RGs, the same technique is now incorporated into the default readout mode options on H4RGs called ``interleaved pixel mode''. Unfortunately at the time of this study, NG has not yet incorporated this readout mode into their simulations. However, the estimated total noise reduction when using IRS2 is about 15\% (B. Rauscher, private communication, 2019). Reducing the results of Figure~\ref{fig:RNresult} proportionally gives $\sigma_{RV}=32~\mathrm{cms}^{-1}$. While this estimate falls short of the target, a more detailed study of the IRS2 may reveal additional benefits. IRS2 not only lowers total read noise levels but also reduces correlated noise and $1/f$ banding\cite{Rauscher_17}.  

\subsection{Pixel Linearity}
\label{sec:linresults}

In this study, the effect of nonlinear pixel response on RV measurements is investigated using the linear-to-nonlinear flux mapping outlined in \S\ref{sec:nonlinear}. Noiseless simulated spectral frames are scaled to integration times between 5 and 60 minutes in order to sample from both the linear nonlinear regions, shown in Figure~\ref{fig:linearitymodel}. RV offsets are derived using the same cross-correlation techniques used throughout this study.

\begin{figure}[ht]\label{fig:linearity}
\begin{center}
\includegraphics[width=0.8\textwidth]{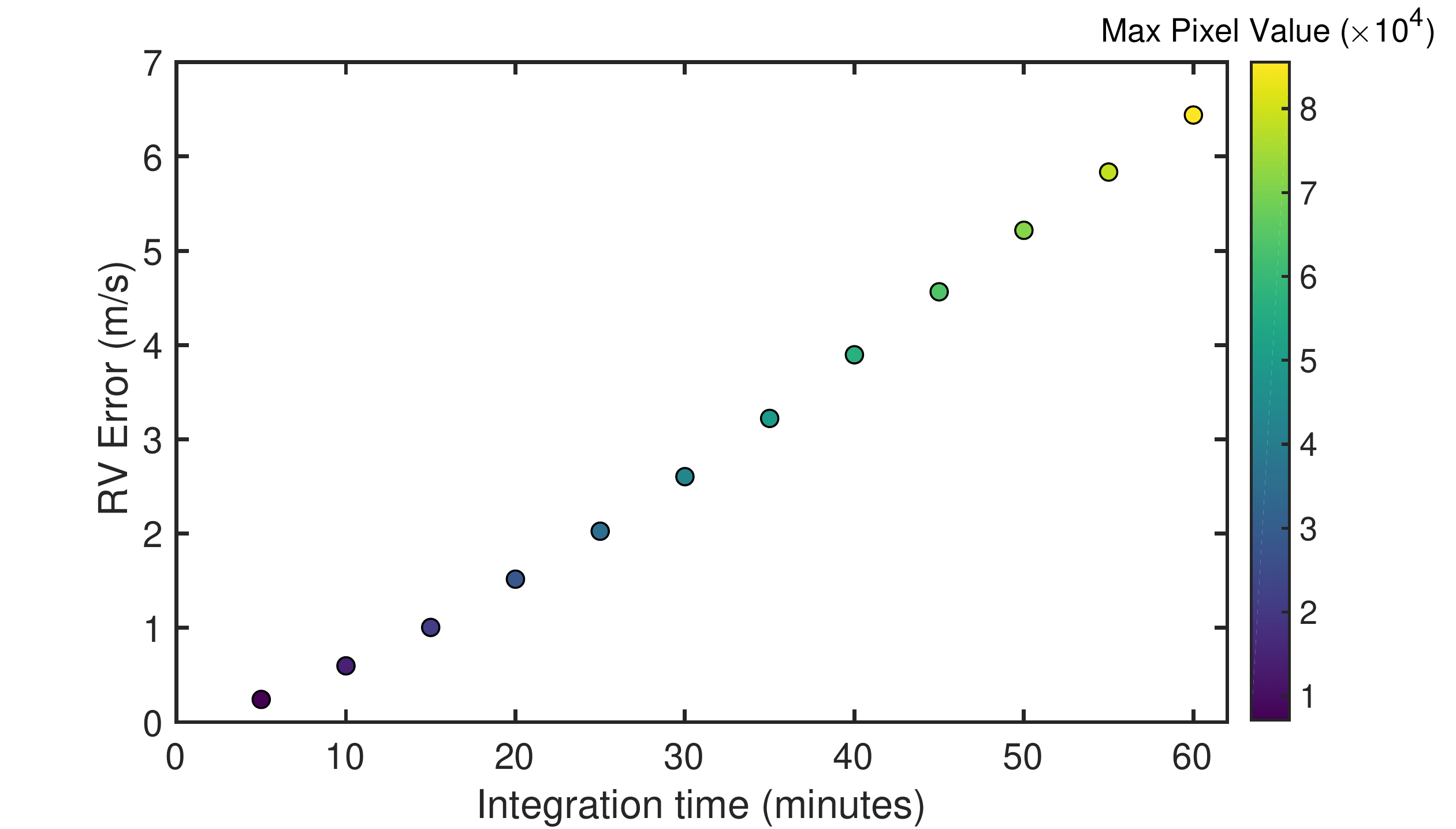}
\end{center}
\caption{The impact of nonlinear pixel response on RV precision. Nonlinearity mapping was derived from Artigau\cite{Artigau_18}. The maximum pixel value colorbar indicates the maximum pixel value of each integrated frame.}
\end{figure}

Nonlinearity results are shown Figure~\ref{fig:linearity}. The relationship between integration time and RV error is nearly linear and, without any correction, even modest integration times result in large RV errors. The maximum pixel value colorbar indicates the maximum count on each frame after integration and also has a linear relationship with RV error. Typical correction techniques involve fitting a series of white-light up-the-ramp samples that integrate beyond typical pixel saturation limits with a low order polynomial. In the case of SPIRou, after applying linearity correction, flux levels were measured to be linear within 0.3\% at a maximum pixel count of $30\times10^4$\cite{Artigau_18}. Applying this correction estimate to the test case shown in Figure~\ref{fig:linearity} at iLocater's nominal integration time of 30 minutes gives an RV error of $0.8$~cms$^{-1}$. Additionally, a new principal component analysis of H4RG detectors has been shown to provide valuable time-domain information that can be useful for further mitigation of intrinsic nonlinear pixel response\cite{Rausher_19}.

\subsection{Dark Current}
\label{sec:darkresults}
Residual dark current is simulated by generating dark frames according to a Poisson distribution, integrating for $\Delta t=30$ minutes and subtracting the mean. A suite of data from a Teledyne study of H2RGs is used to cover temperatures ranging between $T=40-200$K and associated dark current rates. Several additional values from literature have been obtained to explore possible differences in detector manufacturing or environments\cite{Piquette_14,Artigau_18,Blank_12,Zandian_16}. 

\begin{figure*}
\begin{center}
\includegraphics[width=\textwidth]{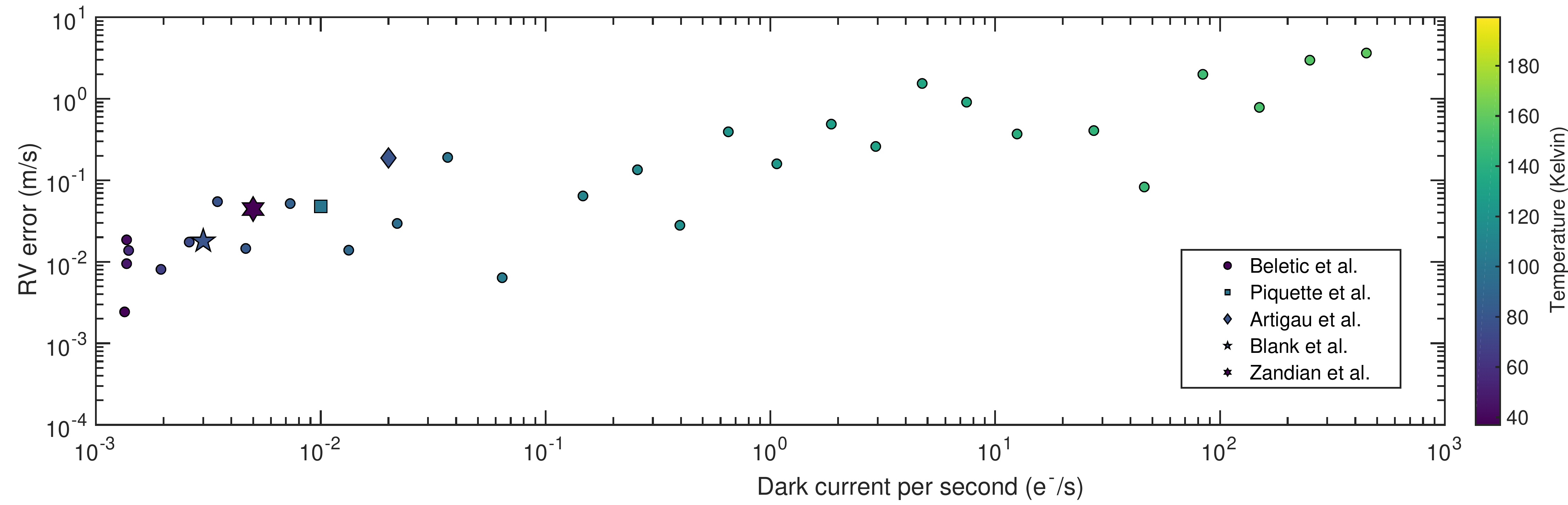}
\end{center}
\caption{\label{fig:dark} The impact of dark current on RV precision. Data was collected from several different sources, with Beletic et al. providing the set marked by circles.\cite{Beletic_08} All data has been rescaled to match the 10 micron pixel pitch of the H4RG-10.}
\end{figure*}

Effective removal of dark current is well documented using standard procedures\cite{McLean_08}. After subtraction, the fixed spatial pattern present in dark current is removed, leaving residual shot noise. Results of dark current simulations are shown in Figure~\ref{fig:dark}. Circles indicate data taken from a study of H2RG detectors by Teledyne\cite{Beletic_08}. Other symbols indicate detectors from the indicated reference. Notably, Artigau et al. 2018 characterize an H4RG detector for SPIRou, an NIR spectrograph operating at 80K. Their dark current findings imply an expected RV impact of $18~\mathrm{cms}^{-1}$ based on our simulations. In fact, all simulated results $T\leq 100$K fall  below $20$~cms$^{-1}$, indicating dark current is not the limiting factor for RV measurements, particularly when compared to read noise.   

\subsection{Persistence}
We measure the effect of persistence on RV by examining three different parasitic, persistent signal scenarios: an M-star with a parasitic M-star; an M-star with a parasitic G-star; and an M-star with parasitic Fabry-Perot etalon calibrator. As the nature of persistent signals results in spatial structure across the detector and because astronomical sources often have very different relative RVs compared to previous targets, it is necessary to maintain the structure of persistent sources and explore the impact of each parasitic source combination at several different relative RV offsets. We consider representative parasitic sources set to 0.01\%, 0.1\%, and 1\% of the primary source. Relative RV offsets between -30~kms$^{-1}$ and 30~kms$^{-1}$ are explored to cover the width of characteristic cross-correlation functional shapes.   

\begin{figure*}[ht]
\begin{center}
\includegraphics[width=\textwidth]{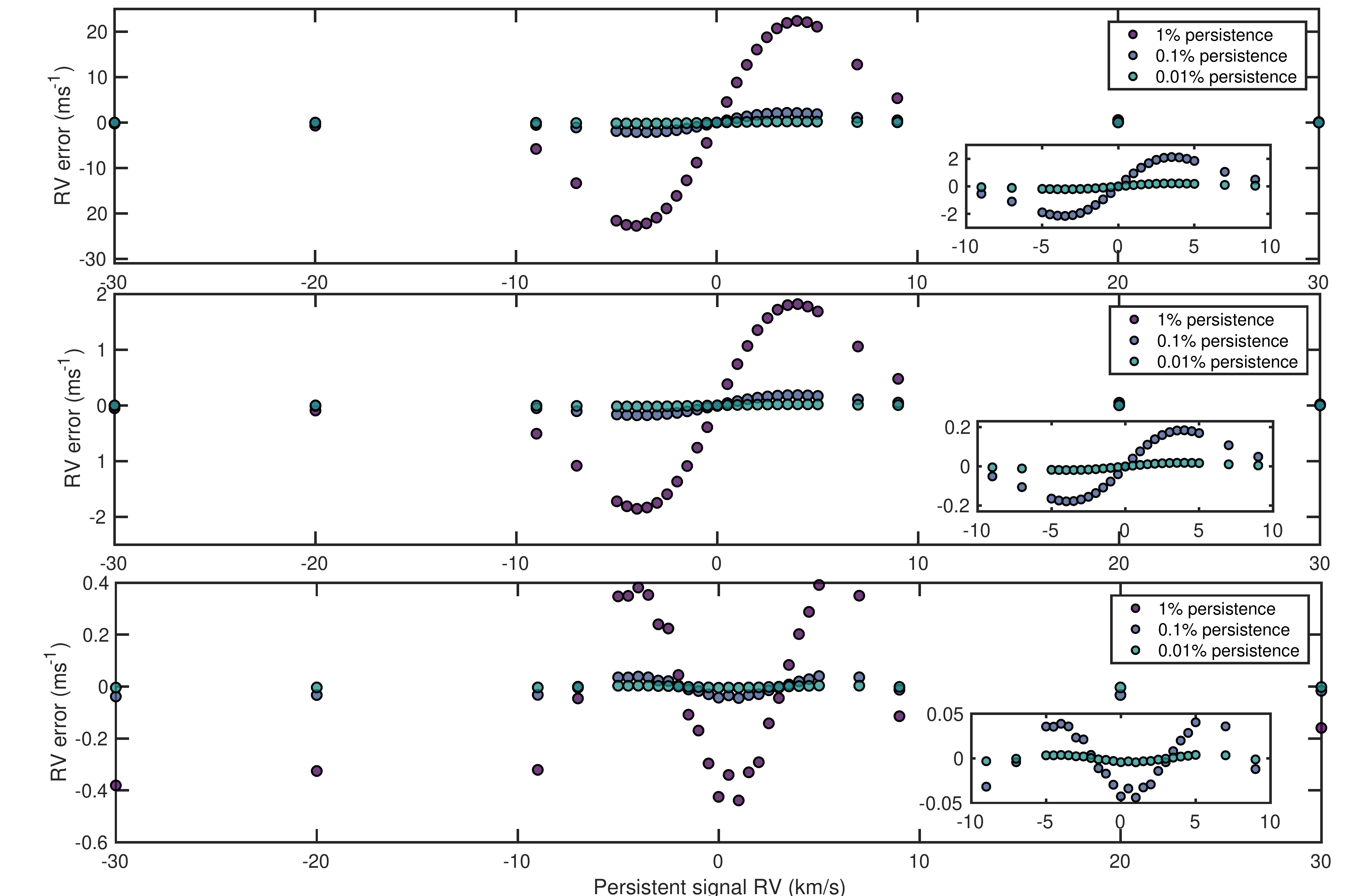}
\end{center}
\caption{\label{fig:allpersist} The impact of differing persistent sources.  \textbf{Top}: M-star with persistent M-star. \textbf{Middle}: M-star with persistent G-star. \textbf{Bottom}: M-star with persistent Fabry-Perot etalon. Persistence fractions are indicated by color. Inset figures show the central portion of the figure in more detail.}
\end{figure*}

Results of simulated persistence for all three cases are shown in the vertical panels of Figure~\ref{fig:allpersist}; insets show the central portion between -10~kms$^{-1}$ and 10~kms$^{-1}$. Persistence fractions are indicated by different colors. Both simulations involving parasitic persistent stellar spectra (top and middle panels) show the same characteristic sinusoidal variation that results from contamination of absorption lines using cross-correlation. Persistent signals with positive relative RV offsets in relation to the primary source affect the average line profile along the ``wings”, positively skewing the velocity fit until the contaminating lines are shifted beyond the primary source line profiles. The same pattern follows for negative relative RVs. RV recovery error reaches a peak when contaminating and primary absorption lines are offset enough to maximally contaminate the edge of the line profile. 

The bottom panel of Figure~\ref{fig:allpersist} shows a similar result when a Fabry-Perot etalon reference is simulated as the parasitic source, which happens during a sequence of calibration exposures bracketing science images. As calibration sources are integrated until very high signal to noise ratios are achieved, realistic persistence fractions are likely to be higher than the previous two scenarios. This suggests observing modes that alternate science frames with calibration frames may not be possible when aiming for sub ms$^{-1}$ precision.    

As persistence decreases with time according to 1/t, the exact ratio of primary source brightness to persistent signal can vary depending on time between exposures, primary source brightness and persistent source brightness\cite{Artigau_18}. While all possible combinations of these parameters represents a vast parameter space, the results shown in Figure~\ref{fig:allpersist} demonstrate thresholds that should be met to minimize RV uncertainty according to a desired instrument precision. Additionally, target similarity and calibration techniques should be carefully considered in the nightly scheduling procedure. For instance, using a dedicated calibration fiber may provide some relief from persistence to prevent calibration light from contaminating the same pixels illuminated by science targets.  

\subsection{Interpixel Capacitance}
Interpixel capacitance is simulated by convolving the detector frame with a kernel, like those described in \S\ref{sec:ipc_theory}. The $3\times3$ kernel with non-zero diagonal components (Equation~\ref{eq:diagIPC}) is generated with
$\alpha$ ranging between $0.01-0.1$ to reasonably explore the parameter space. Measurements performed on WFIRST H4RG-10 detectors indicate $\alpha$ = 0.2 and $\alpha'$ = 0.02 \cite{Kannawadi_16}. Additional IPC kernels tested include a $3\times3$ kernel reported for Hubble's WFC3, an H1RG, and a $9\times9$ kernel measured by SPIRou, an H4RG\cite{Kannawadi_16,Artigau_18}.

\begin{figure*}
    \centering
    \includegraphics[width=\textwidth]{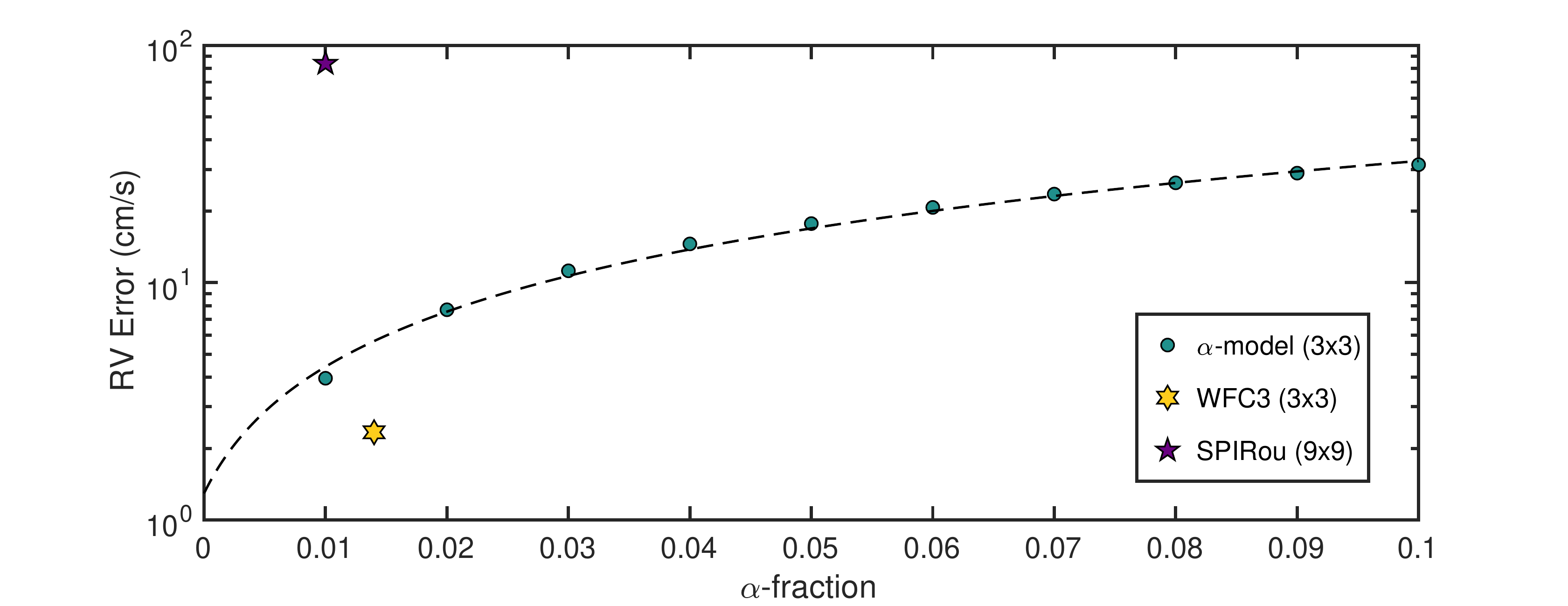}
    \caption{\label{fig:IPCresults} IPC results are shown above. Each circular dot marks the varying fraction of $\alpha$ used in the 3x3 parameterized model, the yellow hexagram and purple star mark experimental measurements of WFC3 and SPIRou`s H4RG. While experimentally measured kernels do not exactly follow the mathematical $\alpha$ model exactly, they are marked at respective average $\alpha$ values.}
\end{figure*}

The results for all tested IPC kernels are shown in Figure~\ref{fig:IPCresults}. As expected, greater $\alpha$ values form a linear relationship with RV error, as the shape and symmetry of the kernel is unchanged. Results for WFC3 also suggest that IPC values in older Teledyne arrays are not necessarily worse than newer detectors, likely due to a reduction in pixel size, reducing overall pixel separation and increasing relative capacitance. SPIRou's kernel likely provides a more realistic result, as the IPC has been measured much farther out than only nearest neighbor pixels. SPIRou's kernel also contains the most asymmetry, which is reflected in its greater impact on RV. Overall, results in Figure~\ref{fig:IPCresults} suggest that RV errors arising from IPC could result in minimal RV errors or limit RV precision to almost $1~\mathrm{ms}^{-1}$, depending on the IPC kernel symmetry.

Mitigating IPC during data processing can be challenging. In the absence of detector noise, a deconvolution kernel, essentially the inverse IPC kernel, can be directly applied to recover the original image and completely remove the effects of IPC. However, as read noise and quantization noise are added after IPC convolution takes place, any attempt at deconvolution would result in more correlated noise\cite{McCullough_08}. While it could be argued that completely static IPC will not inhibit relative RV measurements, as it will be present in each frame in an identical fashion, new research suggests that IPC also has a spatial variability depending on the epoxy voids between the indium bumps that affect capacitance, greatly complicating any attempt at mitigation\cite{Donlon_18}.    


\section{Summary \& Discussion}
\label{sec:concl}

\begin{table}[!t]
	\begin{threeparttable}
		\centerline{
			\begin{tabular}{|l|c|c|c|}
				\hhline{|====|}
				Noise Type & optimistic $\sigma_{RV}~\mathrm{(ms^{-1})}$ & pessimistic $\sigma_{RV}~\mathrm{(ms^{-1})}$ & Notes\\
				\hhline{|=|=|=|=|}
				Read noise & 0.31 & 0.37 & 1\\
				\hline
				Linearity & 0.008 & 2.6 & 2\\
				\hline
				Residual dark current & 0.05 & 0.18 & 3 \\
				\hline	
				Persistence & 0.003 & 2.1 & 4 \\
				\hline
				Interpixel capacitance & 0.05 & 0.87 & 5 \\
				\hline
				Quadrature sum & \textbf{0.32} & \textbf{3.5} & -\\
				\hline
		\end{tabular}}
			\begin{tablenotes}
	 \footnotesize
	 \item \hspace{0.4in}[1] Optimistic values assume a 15\% improvement over results in Figure~\ref{fig:RNresult}.
 	 \item \hspace{0.4in}[2] 0.3\% deviation of pixel response from linear used for 
 	 \item \hspace{0.55in} optimistic estimate at 30 minutes integration time.
	 \item \hspace{0.4in}[3] $T=80$K nominal operating temperature.
	 \item \hspace{0.4in}[4] Optimistic values computed using best-case observing scenarios and   
	 \item \hspace{0.55in} averaging results beyond the clear envelope.
	 \item \hspace{0.4in}[5] $\alpha$-model H4RG values and SPIRou's kernel results are selected for  
	 \item \hspace{0.55in} optimistic and pessimistic models, respectively.
	 
		\end{tablenotes}
	\caption[Summary of Results.]{\label{tab:SummaryOfResults}Summary of detector effects. As a number of the detector effects listed are based on a statistical or random manifestation of noise, expected values for each will likely lie somewhere between the upper and lower limits noted in the optimistic and pessimistic columns. Summing each column in quadrature gives an approximate maximum RV error.}
		\
	\end{threeparttable}
\end{table}

New H4RG detectors contain four times as many contiguous pixels as the previous generation of infrared arrays and have quantum efficiencies that exceed $90\%$\cite{Zandian_16}. Accordingly, the H4RG detector may be well-suited for cross-dispersed, echelle spectrographs that operate at high order number. While the H4RG has attractive qualities, its noise characteristics, and how they relate to RV measurements, have not previously been explored in detail. 

We have presented the first investigation into the impact of H4RG detectors on RV precision using simulations of a high-resolution spectrograph and noise generator that emulates Teledyne's HxRG arrays. Read noise, pixel nonlinearity, dark current, persistence, and IPC were studied, and RV uncertainties quantified for each source. Table~\ref{tab:SummaryOfResults} shows a summary for each detector effect, assuming an optical model that resembles the iLocater spectrograph as a representative case. For the majority of effects analyzed, it is difficult to place a single RV value, considering the many dependant factors involved. Therefore, we have chosen to present optimistic and pessimistic estimates to provide an expected range for each, taking into account possible improvements in mitigation, e.g. new readout schemes and careful target scheduling. 

Optimistic estimates for dark current, persistence, and IPC are chosen by selecting the minimum values from a given study. In the case of read noise, IRS2 is expected to provide improvements of about $15\%$, and for linearity effects, SPIRou has shown a correction level below 1\%. Pessimistic values for read noise, dark current, and IPC are selected in the contrary scenario; maximum values are selected from each study. For persistence, pessimistic estimates assume a $0.1\%$ remnant image of an identical spectral type at the most detrimental RV offset of $RV =3.5~\mathrm{kms}^{-1}$ leading to an RV error of $\sigma_{RV}=2.1~\mathrm{ms}^{-1}$. No linearity correction is applied for the pessimistic linearity case, using iLocater's nominal integration time of 30 minutes. 

Persistence and linearity effects pose the greatest risk, by a factor of several compared to the other effects studied, and could easily dominate the entire RV error budget if left unchecked. Assuming appropriate observing procedures are in place to deal with persistence and typical linearity correction is applied, the quadrature sum of read noise, pixel nonlinearity, dark current, and IPC results in the range $\sigma_{RV}=0.32-3.5~\mathrm{ms}^{-1}$. A series of tests that included all detector effects indicates in many cases the errors in Table~\ref{tab:SummaryOfResults} can result in partial cancellation, whereby some individual effects may result in a net blue-shift while others a net red-shift. As such, summing errors in quadrature might be considered an upper limit. These results suggest that precision RV spectrographs could generate sub-meter-per-second precision in the near-infrared if HxRG arrays are used and error budgets are dominated by detector noise. 

The simulations performed in this paper are capable of reliably recreating most of the noise characteristics of HxRGs, however they do not encompass every source of detector noise. Particularly, a few of the more unusual noise characteristics are not included, e.g. random telegraph noise, intra-pixel QE variations and the brighter-fatter effect.\cite{shapiro_18,choi_19} However, recently published research by Rauscher demonstrates a new technique to preserve time domain information from HxRG datacubes which could be used to mitigate a number of usual and unusual detector effects including pixel nonlinearity, persistence, and the brighter-fatter effect\cite{Rausher_19}.

While simulations of H4RGs is work that is still on-going, the most practical next step in characterizing performance would involve laboratory experiments to validate the errors in Table~\ref{tab:SummaryOfResults}. Beyond a first order treatment, it is likely that detector noise characteristics are batch-specific if not detector-specific within a batch. While there is certainly still room for improvement, H4RGs appear to be suitable for Doppler spectrometers aiming for sub ms$^{-1}$ precision. Any further technical advances made by the WFIRST mission will directly benefit infrared spectrographs like SPIRou, iLocater, and others that baseline the H4RG large format detector. 

\section{Acknowledgements}
We thank Bernard Rauscher for many useful discussions on detector systems and his guidance in integrating NG into our simulator environment. Justin R. Crepp acknowledges support from the NASA Early Career and NSF CAREER Fellowship programs. 
\bibliography{article.bib}   
\bibliographystyle{spiejour.bst}   
\listoffigures

\end{spacing}
\end{document}